\documentclass[a4paper,fleqn]{cas-dc}

\usepackage{graphicx}
\usepackage{microtype} 
\usepackage[numbers]{natbib}

\usepackage{float} 
\usepackage{stfloats} 
\usepackage{booktabs} 
\usepackage{array} 

\def\tsc#1{\csdef{#1}{\textsc{\lowercase{#1}}\xspace}}
\tsc{WGM}
\tsc{QE}
\tsc{EP}
\tsc{PMS}
\tsc{BEC}
\tsc{DE}

\begin{document}
\let\WriteBookmarks\relax
\def\floatpagepagefraction{1}
\def\textpagefraction{.001}

\shorttitle{U-Net in Medical Image Segmentation}

\shortauthors{neha et~al.}

\title [mode = title]{U-Net in Medical Image Segmentation: A Review of Its Applications Across Modalities}                      


%
\author[1]{Fnu Neha}[orcid=0009-0004-3702-2382]

\cormark[2]

\ead{neha@kent.edu}



\affiliation[1]{organization={Department of Computer Science, Kent State University},
    city={Kent},
    state={Ohio},
    country={USA}}

\author[1]{Deepshikha Bhati} [orcid=0009-0002-0115-6026]
\ead{dbhati@kent.edu}


\author[2]{Deepak Kumar Shukla}[orcid = 0009-0008-3813-4959]
\ead{ds1640@scarletmail.rutgers.edu}


\affiliation[2]{organization={Rutgers Business School, Rutgers University},
    city={Newark},
    state={New Jersey},
    country={USA}}
    
\author[1]{Sonavi Makarand Dalvi} [orcid = 0009-0005-7037-3229]
\cormark[1]
\ead{sdalvi@kent.edu}


\author[3]{Nikolaos Mantzou}[orcid = 0009-0003-7870-5633]
\ead{nikolaosmantzou@gmail.com}

\affiliation[3]{organization={School of Medicine, Aristotle University of Thessaloniki},
    state={Thessaloniki},
    country={Greece}}

\author[1]{Safa Shubbar} [orcid = 0000-0002-0244-5391]
\ead{sshubbar@kent.edu}


\cortext[cor1]{Corresponding Author}
\cortext[cor2]{Principal Corresponding Author}



\begin{abstract}
Medical imaging is essential in healthcare to provide key insights into patient anatomy and pathology, aiding in diagnosis and treatment. Non-invasive techniques such as X-ray, Magnetic Resonance Imaging (MRI), Computed Tomography (CT), and Ultrasound (US), capture detailed images of organs, tissues, and abnormalities. Effective analysis of these images requires precise segmentation to delineate regions of interest (ROI), such as organs or lesions. Traditional segmentation methods, relying on manual feature-extraction, are labor-intensive and vary across experts. Recent advancements in Artificial Intelligence (AI) and Deep Learning (DL), particularly convolutional models such as U-Net and its variants (U-Net++ and U-Net 3+), have transformed medical image segmentation (MIS) by automating the process and enhancing accuracy. These models enable efficient, precise pixel-wise classification across various imaging modalities, overcoming the limitations of manual segmentation. This review explores various medical imaging techniques, examines the U-Net architectures and their adaptations, and discusses their application across different modalities. It also identifies common challenges in MIS and proposes potential solutions.
\end{abstract}


\begin{highlights}
\item  A detailed discussion of X-ray, MRI, CT, and US, highlighting their types and importance in healthcare.
\item A comprehensive overview of U-Net, tracing its development and recent architectural advancements.
\item An analysis of U-Net's applications across different medical images.
\item A discussion on the limitations in current techniques (architecture and modality) and suggestions for future advancements in the field.
\end{highlights}


\maketitle

\section{Introduction} \label{sec1}

Images serve as critical tools for capturing and conveying information about objects, scenes, or concepts through various techniques such as photography, digital imaging, or specialized sensing technologies. In healthcare, medical imaging (MI) holds a pivotal role in both diagnosis and treatment. These images are acquired using non-invasive imaging modalities, including X-rays, Magnetic Resonance Imaging (MRI), Computed Tomography (CT), and Ultrasound (US). Each modality offers distinct advantages and is tailored for specific diagnostic applications, enabling precise visualization and analysis of anatomical structures and physiological processes \cite{hussain2022modern, bercovich2018medical}. For example, MRI scans are invaluable for assessing brain tumors, and CT scans are used to evaluate internal injuries \cite{ellingson2014pros, hussain2022modern}. In contrast to general-purpose imagery, MI demands advanced techniques to capture detailed anatomical and pathological features. These methods are essential for ensuring accurate visualization, interpretation, and analysis, which are critical for effective clinical decision-making and treatment planning.

One important aspect of MI is segmentation, which involves identifying and delineating regions of interest (ROI), such as organs or lesions \cite{rayed2024deep}. This process is essential for extracting key information about the shape, size, and volume of these structures. Traditional segmentation methods rely on manual feature-extraction and techniques like thresholding or edge detection \cite{yu2023techniques}. These methods are time-consuming and inefficient, and these are subject to variability due to reliance on expert input. This inconsistency has driven the demand for more advanced, automated methods to enhance the segmentation process.

The integration of advanced computational techniques has significantly enhanced MI by enabling the automated interpretation and analysis of complex image data. These innovations streamline diagnostic processes and improve the accuracy of image-based assessments \cite{chauhan2024unleashing}. Deep learning (DL), a subset of AI, uses neural networks with multiple layers to learn patterns from vast datasets. These networks automate tasks which previously required specialized human expertise, such as detecting anomalies in medical images or predicting patient outcomes.

Recent advancements in DL for Medical Image Segmentation (MIS) have established automatic segmentation as a superior alternative to traditional methods \cite{swarnkar2022deep, trimpl2022beyond, zi2024research}. The popular segmentation techniques are: (1) semantic segmentation: classifies each pixel \cite{zi2024research}; (2) instance segmentation: identifies individual objects \cite{hafiz2020survey}; (3) panoptic segmentation: combines both semantic and instance segmentation, provide a comprehensive understanding of the image. Panoptic segmentation assigns a label to each pixel while differentiating between instances of objects within the same category \cite{kirillov2019panoptic}. These techniques have achieved remarkable accuracy across various medical imaging datasets, including the International Skin Imaging Collaboration (ISIC) for skin cancer \cite{8363547}, Brain Tumor Segmentation (BraTS) for brain tumors \cite{ghaffari2019automated}, and Kidney Tumor Segmentation (KiTS) for kidney tumors \cite{heller2019kits19}.

As image-based diagnosis becomes increasingly crucial in clinical practice, encoder-decoder models like U-Net and its variants \cite{ronneberger2015u, zhou2018unet++, huang2020unet} have gained prominence for their flexibility and strong performance in MIS. The growing emphasis on deep learning (DL) has led to a significant body of research and review articles on MIS, underscoring the pivotal role of DL models in advancing efficient computer-aided diagnosis systems. These evolving models hold the promise of making diagnoses faster, more accurate, and more accessible.

While existing reviews, such as those by Siddique et al. \cite{siddique2021u} and Azad et al. \cite{azad2024medical}, provide comprehensive overviews of U-Net and its architectural variants across various medical imaging modalities, their focus has primarily been on theoretical advancements and applications. These studies explore the development of U-Net and its use in MIS, with an emphasis on its role in improving diagnostic accuracy. In contrast, our review includes the latest developments in the field and expands on recent advancements in both U-Net architectures and the diverse modalities in which they are applied.

Our paper extends existing reviews by focusing on the practical integration of U-Net in clinical settings, particularly emphasizing the role of semantic segmentation techniques in improving diagnostic accuracy and treatment planning. We further enhance model interpretability by incorporating predefined semantic features, addressing a key gap between advanced segmentation methods and their real-world healthcare applications. To provide a comprehensive and up-to-date perspective on U-Net-based methods, we include recent review papers and explore a broader spectrum of medical imaging modalities. For each modality, we have included detailed attributes such as Study, Modality, Focus Area, Methodology, and Performance Metrics, offering valuable insights into their specific applications and contributions to medical image segmentation.

The key contributions of our paper include:
\begin{enumerate} 
\item \textbf{Overview of Medical Imaging Modalities:} A comprehensive exploration of X-ray, MRI, CT, and Ultrasound modalities, focusing on their classifications, applications, and critical roles in medical diagnostics and healthcare. 
\item \textbf{U-Net and Its Variants:} A thorough review of the U-Net architecture, tracing its development, variants, and recent advancements. 
\item \textbf{Applications Across Modalities:} An analysis of U-Net's diverse applications across various medical imaging types, with detailed attributes such as Study, Modality, Focus Area, Methodology, and Performance Metrics for each modality. 
\item \textbf{Limitations and Future Directions:} A discussion of the current limitations in both architectural approaches and modality-specific applications, along with suggestions for future advancements in the field. 
\end{enumerate}

The paper is organized as follows: Section \ref{sec2} presents the research methodology. Section \ref{sec3} provides an overview of the common medical imaging modalities. Section \ref{sec4} discusses U-Net and its variants. Section \ref{sec5} reviews U-Net's applications across medical images. Section \ref{sec6} identifies limitations. Section \ref{sec7} presents the discussion and future directions. Section \ref{sec8} concludes the review.

\section{Research Methodology}\label{sec2}

This review explores the application of U-Net in healthcare, focusing on its use across various medical imaging modalities to provide a comprehensive overview. Our research methodology is outlined as follows:

\begin{itemize}
    \item \textbf{Literature Search:} A thorough search was conducted across electronic databases, including \emph{PubMed, Elsevier, Scopus, Google Scholar} and \emph{IEEE Xplore}. Top medical imaging conferences, such as \emph{Medical Image Computing and Computer-Assisted Intervention (MICCAI), International Symposium on Biomedical Imaging (ISBI) ,} and \emph{Information Processing in Medical Imaging (IPMI)}, were also included to gather relevant materials. The search strategy involved keywords related to \emph{U-Net, medical imaging, healthcare applications,} and \emph{deep learning} to ensure comprehensive coverage.
    
    \item \textbf{Inclusion and Exclusion Criteria:}
    \begin{itemize}
        \item \textbf{Inclusion:} Studies specifically addressing U-Net's implementation in healthcare contexts, including segmentation, classification, and diagnosis.
        \item \textbf{Exclusion:} Articles not published in English, those lacking full-text availability, and studies unrelated to U-Net or healthcare/medical imaging applications.
    \end{itemize}
    
    \item \textbf{Data Retrieval and Screening:} As of October 2024, approximately 150 recent publications were retrieved and screened for relevance based on the predefined criteria.
    
    \item \textbf{Framework for Literature Categorization} A framework was developed to categorize the literature by medical imaging modalities (e.g., x-ray, magnetic resonance imaging, computed tomography and ultrasound scans), highlighting U-Net's versatility across modalities.

    \item \textbf{Synthesis and Analysis:} Findings were synthesized to summarize U-Net's contributions to diagnostic accuracy. The analysis discussed limitations and potential improvements, paving the way for future directions.
\end{itemize}

\subsection{Research Objective}

The primary research objective of this review is to understand, how U-Net and various imaging modalities integrate to manage organ-related diseases, enhancing the effectiveness of diagnostics, treatment recommendations, and overall patient health.

\section{Medical Imaging Modalities} \label{sec3}

In this section, we will cover key concepts related to various medical imaging modalities or techniques, including X-ray, MRI, CT, and Ultrasound.

\subsection{X-Ray}

\begin{figure}
    \centering
    \includegraphics[width=1\linewidth]{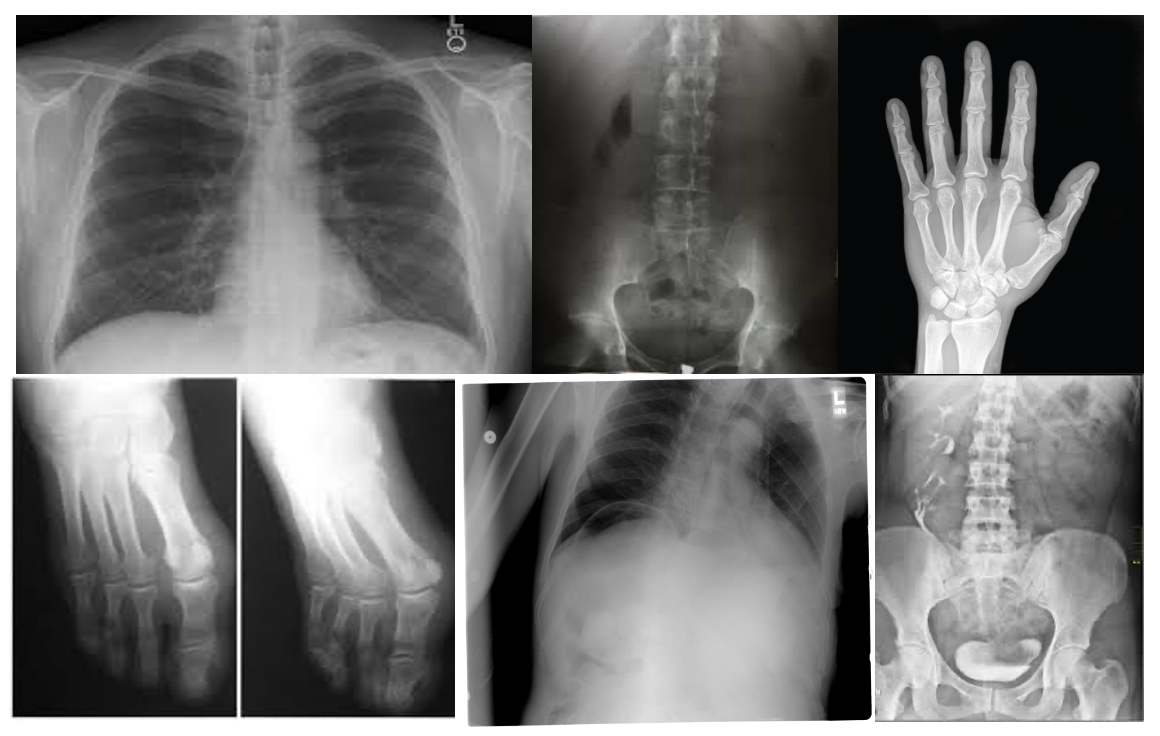}
    \caption{\textls[-15]{X-Ray - a non-invasive diagnostic medical imaging modality taken from \cite{wiki:xxx,wiki:xxx1, wiki:xxx2, wiki:xxx3,wiki:xxx4, wiki:xxx5}}}
    \label{fig:xray}
\end{figure}

X-rays are a form of electromagnetic radiation, similar to visible light, with much higher energy and shorter wavelengths \cite{hessenbruch2002brief, prabhu2020production, hoheisel2006review, brown2012x}. These are essential in the medical field for diagnostic imaging, enabling healthcare professionals to non-invasively visualize the body's internal structures. This technology allows for detailed imaging of bones, organs, and tissues, helping diagnose injuries, detect diseases, and guide treatment decisions accurately and efficiently as shown in Fig. \ref{fig:xray}.

These are produced in an X-ray tube, consisting of a cathode (the negative electrode) and an anode (the positive electrode). When the cathode is heated, it emits electrons, which are then accelerated toward the anode under the influence of a high-voltage potential, typically ranging from 20 to 100 kV. When the high-speed electrons strike the anode, they undergo two processes:
\begin{itemize}
    \item \textbf{Bremsstrahlung Radiation:} The high-energy electrons are decelerated upon interaction with the positive electric field of the anode nucleus, leading to the release of X-ray photons as a result of the energy loss during deceleration. 
    \item \textbf{Characteristic Radiation:} When electrons possess sufficient energy, they can eject inner-shell electrons from the atoms of the anode material, typically tungsten. This process creates vacancies that are subsequently filled by outer-shell electrons transitioning to lower energy states, resulting in the emission of characteristic X-rays.
\end{itemize}

\subsubsection{Properties of X-rays}
\begin{itemize}
    \item \textbf{Penetrating Power:} X-rays can penetrate body tissues and degree of penetration depends on the energy of the X-rays and the density of the material they pass through.
    \item \textbf{Detection:} X-rays can be detected using film (traditional radiography) or digital detectors (computed radiography or direct digital radiography). These detectors convert X-rays into images by capturing the varying levels of radiation transmitted through the body.
\end{itemize}

\subsubsection{Image Formation}
When X-rays pass through the body, their absorption varies depending on the density and composition of the tissues:
\begin{itemize}
    \item \textbf{Dense Structures:} High-density structures, such as bones, absorb a substantial proportion of X-rays, resulting in their white appearance on radiographic images.
    \item \textbf{Soft Tissues:} Structures like organs and muscles exhibit moderate absorption, appearing as varying shades of gray.
    \item \textbf{Air:} Low-density regions, such as those containing air (e.g., the lungs), absorb minimal X-rays, leading to a dark or black representation on the image.
\end{itemize}

\subsection{Magnetic Resonance Imaging (MRI)}

Magnetic Resonance Imaging (MRI) is a non-invasive medical imaging technique to provide high-resolution images of internal organs and tissues \cite{khoo1997magnetic, manduca2001magnetic, katti2011magnetic, vlaardingerbroek2013magnetic}. MRI differentiates itself from X-ray imaging by avoiding the use of ionizing radiation. Instead, MRI works by aligning hydrogen protons in the body using a strong electromagnetic field. These protons are then disrupted by a radiofrequency (RF) pulse, and as they return to their original alignment, they emit energy detected by the scanner to produce detailed images. Its key points are: 
\begin{itemize}
    \item \textbf{Magnetic Field (MF)}: The MRI machine consists of a large magnet generates a strong MF, ranging from 0.5 and 1.5 Tesla. This MF aligns the protons in the hydrogen atoms of the body's tissues.
    \item \textbf{Radiofrequency (RF) Pulses}: After the alignment of protons within the MF, RF pulses are applied to the targeted region. These RF pulses momentarily disturb the equilibrium alignment of the protons, causing them to shift from their original orientation.
    \item \textbf{Relaxation}: Following the cessation of the RF pulses, the protons gradually realign with the main MF. During this realignment process, they emit energy in the form of RF signals. This phenomenon, referred to as relaxation, varies in rate depending on the specific properties of the tissue being imaged.
    \item \textbf{Signal Detection}: The emitted radio waves are detected by the MRI machine and converted into electrical signals. These signals are then processed by a computer to create images of the scanned area.
\end{itemize}

MRI is particularly effective for imaging soft tissue and nervous tissue, making it ideal for assessing injuries to cartilaginous structures and ligaments (e.g., ankle or cruciate ligament injuries), evaluating tumors (e.g., breast cancer), and diagnosing central nervous system disorders (e.g., encephalitis, demyelination, acoustic neuroma).

\subsection{Types of MRI}
\begin{figure}
    \centering
    \includegraphics[width=1\linewidth]{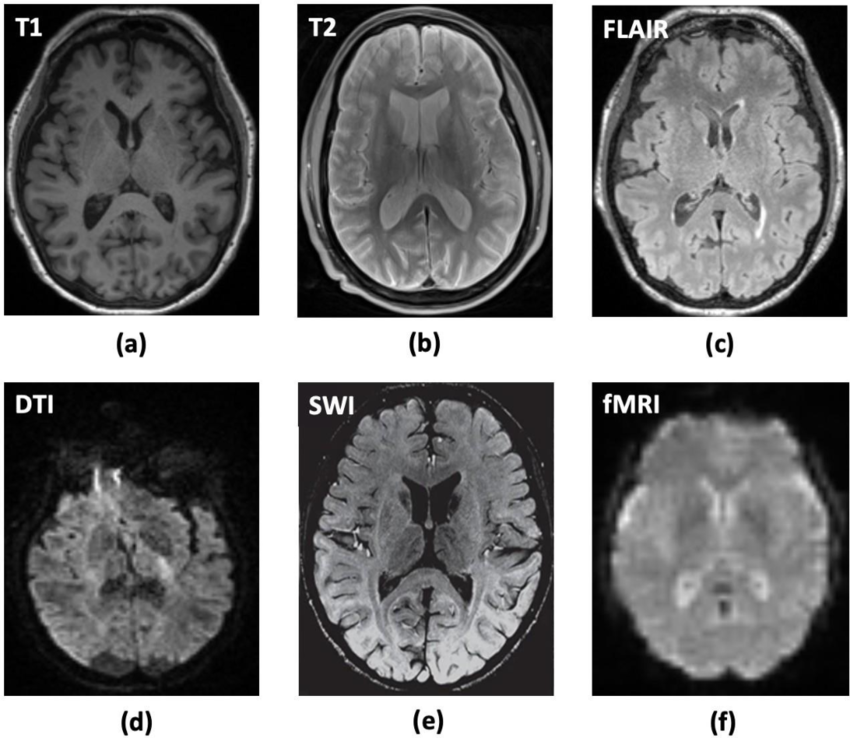}
    \caption{\textls[-15]{Brain MRI - a non-invasive high-resolution medical imaging modality taken from \cite{gu2023exploring} Fig. 2(a) T1-weighted MRI, 2(b) T2-weighted MRI, 2(c) Fluid-Attenuated Inversion Recovery (FLAIR), 2(d) Diffusion Tensor Imaging (DTI), 2(e) Susceptibility Weighted Imaging (SWI), 2(f) Functional MRI (fMRI)}}
    \label{fig:mri}
\end{figure}

MRI can be categorized into several types, each designed to provide specific information about the body’s structures and functions as shown in Fig. \ref{fig:mri}. 

\begin{itemize}
    \item \textbf{Structural MRI}:
    \begin{itemize}
        \item \textbf{T1-weighted MRI}: These images provide high-resolution anatomical detail \cite{caramella2019essentials}. T1-weighted imaging sequences are particularly effective for highlighting fat-rich tissues and evaluating normal anatomical structures. These sequences generate high-intensity signals for tissues with high fat content, while fluid-filled structures appear with lower signal intensity, providing distinct contrast for diagnostic assessment.
        \item \textbf{T2-weighted MRI}: T2-weighted imaging is highly sensitive to variations in water content within tissues, making it a valuable tool for detecting pathological conditions such as tumors, inflammatory processes, and edema.  \cite{caramella2019essentials}. These images enhance the contrast of fluid-rich regions, aiding in precise diagnostic evaluations.

    \end{itemize}

    \item \textbf{Fluid-Attenuated Inversion Recovery (FLAIR)}: 
    FLAIR is a specialized MRI sequence that suppresses the signal from cerebrospinal fluid (CSF), making it easier to visualize lesions and abnormalities in the brain \cite{kates1996fluid}. It is particularly useful for detecting white matter lesions.

    \item \textbf{Diffusion MRI}: This technique evaluates the diffusion of water molecules within tissues \cite{jones2010diffusion}. It is particularly useful for imaging brain white matter tracts, as it provides insights into the integrity of neural pathways.
    \begin{itemize}
        \item \textbf{Diffusion Tensor Imaging (DTI)}: A specialized form of diffusion MRI that characterizes the directionality of water diffusion \cite{le2001diffusion}. DTI allows for the visualization of white matter tracts, which is essential in studying conditions like stroke and multiple sclerosis.

    \end{itemize}
    
    \item \textbf{Susceptibility Weighted Imaging (SWI)}: 
    SWI is an advanced MRI technique that enhances the visualization of blood vessels and detects small hemorrhages by utilizing phase information from the MR signal \cite{haller2021susceptibility}. It is particularly valuable in identifying vascular malformations and assessing traumatic brain injuries.
    
    \item \textbf{Functional MRI (fMRI)}: fMRI measures brain activity by detecting changes in blood flow related to neural activity \cite{deyoe1994functional}. When a brain region is active, it consumes more oxygen, leading to changes in the blood’s oxygenation level. fMRI can be used for pre-surgical brain mapping, studying brain functions, and assessing neurological disorders.

\end{itemize}

\subsection{Computed Tomography (CT)}

Computed Tomography (CT) scans use rotating X-ray generators to create detailed cross-sectional images of the body \cite{brooks1993computed, gharieb2022computed, cierniak2011x}. These rely on the differential absorption of X-rays by various tissues. A CT scanner uses a rotating assembly comprising an X-ray tube and detectors to acquire X-ray projections from multiple angular perspectives. These projections are computationally reconstructed to produce detailed two-dimensional or three-dimensional visualizations of internal anatomical structures. Unlike traditional radiography, which superimposes structures, CT scans produce detailed slices of the subject, often just a few millimeters thick, enabling three-dimensional reconstruction and improved visualization of the said internal structures. CT assigns density-based values called Hounsfield units to voxels, facilitating differentiation of tissues. Contrast-enhanced scans further improve visualization, particularly for blood vessels.

The Hounsfield scale is a measurement system used in CT imaging to assess the radiodensity of various materials. It assigns values called Hounsfield units (HU), with water being the reference point at 0 HU. High-density materials, such as bone, are assigned positive HU values, generally ranging from 1000 to 1500 HU, while low-density substances like air are given negative values, around -1000 HU. The radiodensity levels are depicted in grayscale on CT images, where denser structures appear brighter, and less dense structures appear darker. A voxel is a fundamental three-dimensional unit that forms part of the reconstructed image. Smaller voxels contribute to greater image clarity and detail. Tissues that absorb more x-rays, or have higher attenuation, produce bright voxels, while those with lower absorption result in darker voxels. This differentiation is essential for accurately visualizing tissue structures.

\subsubsection{Image Planes: Axial, Coronal, and Sagittal}
\begin{figure}
    \centering
    \includegraphics[width=1\linewidth]{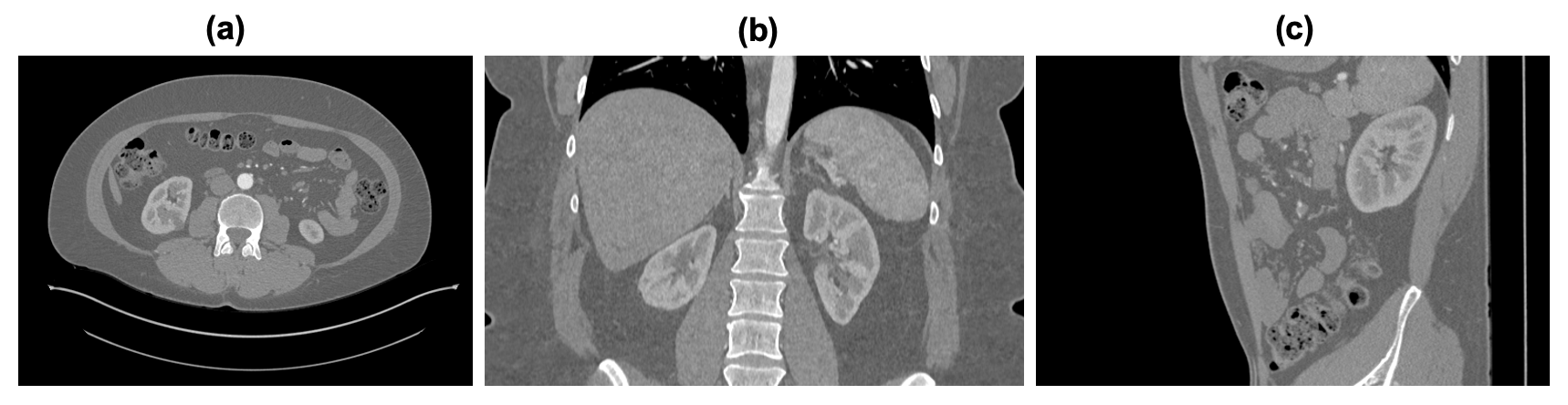}
     \caption{\textls[-15]{Kidney CT scan taken from \cite{heller2019kits19} - Fig. 3(a) axial, 3(b) coronal, and 3(c) sagittal plane }}
    \label{fig:ct}
\end{figure}

CT images can be acquired in multiple planes as shown in Fig. \ref{fig:ct}, including axial, coronal, and sagittal views, which provide different perspectives of the body:
    \begin{itemize}
        \item \textbf{Axial (Transverse) Plane}: This is the most common orientation, providing cross-sectional images of the body in horizontal slices \cite{heller2019kits19}. Each slice corresponds to a specific thickness, typically ranging from 1 to 10 mm, allowing for detailed examination of organs and structures.
        \item \textbf{Coronal Plane}: This orientation provides images viewed from the front, slicing the body into anterior and posterior sections \cite{heller2019kits19}. Coronal views are particularly useful for visualizing structures such as the sinuses, heart, and lungs.
        \item \textbf{Sagittal Plane}: These images divide the body into left and right sections, offering insights into the midline structures \cite{heller2019kits19}. Sagittal views help assess spinal alignment and certain anatomical relationships.

\end{itemize}

\subsection{Types of CT Scans and Phases}
\begin{figure}
    \centering
    \includegraphics[width=1\linewidth]{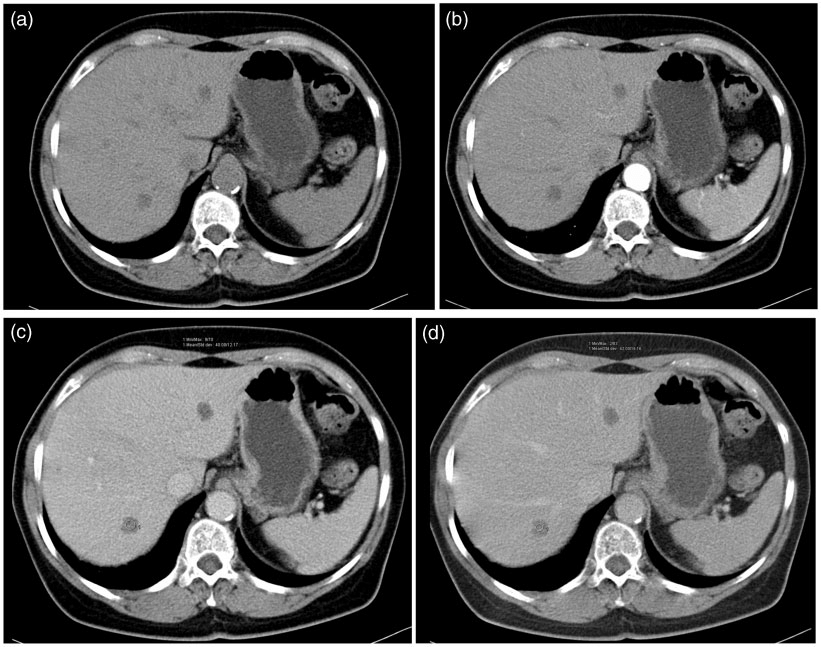}
    \caption{\textls[-15]{Liver CT scan taken from \cite{young2015benign}. Fig. 4(a) Non-Contrast CT, 4(b) Contrast Enchanced CT - Arterial Phase, 4(c) Venous Phase, 4(d) Delayed Phase}}
    \label{fig:liverct}
\end{figure}

CT scans are classified based on their applications and imaging phases as shown in Fig.\ref{fig:liverct}:

\begin{itemize}
    \item \textbf{Non-Contrast CT}: This imaging modality, performed without the administration of a contrast agent, is commonly employed as an initial diagnostic tool for evaluating conditions such as fractures, hemorrhages, and neoplasms \cite{heller2019kits19}. It provides \textbf{essential baseline data regarding tissue density and structural integrity. }

    \item \textbf{Contrast-Enhanced CT}: This scan has an iodine-based contrast agent administered intravenously or orally to enhance the visibility of vascular structures and organs \cite{heller2019kits19}. It is essential in oncological assessments, abdominal imaging, and vascular studies. Following are the phases of Contrast-Enhanced CT:

    \begin{itemize}
        
            \item \textbf{Arterial Phase}: Images are acquired 25-30 seconds after contrast injection, ensuring enhanced visualization of arterial vessels and hypervascular lesions due to their increased contrast enhancement during this phase.
            
            \item \textbf{Venous Phase}: Images are acquired 60-90 seconds post-injection, this phase focuses on venous structures and provides valuable information about tumor perfusion and vascular integrity.
            
            \item \textbf{Delayed Phase}: Captured 10-15 minutes after contrast administration, this phase assesses the distribution of contrast within tissues, particularly beneficial for evaluating renal function and identifying tumors with varying vascularity.

    \end{itemize}
\end{itemize}

\subsection{Ultrasound(US)}
An ultrasound scan, also known as sonography or ultrasonography, utilizes high-frequency sound waves to generate real-time images of structures inside the body as shown in Fig. \ref{fig:us} \cite{debdas2014scientific, szabo2013diagnostic, royer2019seeing}. US is widely used for diagnostic purposes, offering a non-invasive, safe, and relatively inexpensive method to visualize tissues, organs, and blood flow.

\subsubsection{Working Principle of Ultrasound}
Ultrasound imaging operates on the principle of acoustic reflection, where sound waves are transmitted from a transducer into the body and reflected back from tissues, providing data for image formation \cite{metzner2016ultrasound, chan2011basics, kremkau2015sonography}. The sound waves propagate through various tissues, and upon encountering interfaces between different tissue types, such as muscle and bone or fluid and soft tissue, they are reflected back toward the transducer. These soundwaves, generated and received by piezoelectric transducers in a probe, create images by interpreting echoes. Denser materials produce stronger echoes, appearing brighter on the image, while the time taken for echoes to return indicates the depth of the structure.

Ultrasound can be employed endoscopically to assess difficult-to-reach organs like the prostate, ovaries, pancreas, and heart valves. The transducer functions both as a transmitter, emitting sound waves, and as a receiver, capturing the returning echoes.
The sound waves are typically in the range of 2–20 MHz, above the human hearing range.
A gel is applied between the transducer and the skin to eliminate air gaps, which could otherwise interfere with the transmission of sound waves by causing premature reflection. The modality includes various imaging modes, such as A-mode (plots echoes as peaks reflecting depth), B-mode (produces grayscale, two-dimensional images showing depth and density), M-mode (captures motion sequences of structures like the heart), and Doppler or duplex mode (assesses blood flow velocity and direction using frequency shifts).
The returning echoes are analyzed based on the time delay (how long the echo takes to return) and intensity (strength of the echo) to create a 2D or 3D image on the US monitor.

\subsubsection{Types of US Scans}
US scan has the following types:
\begin{itemize}
    \item \textbf{Endoscopic (EUS)} combines endoscopy and US to obtain high-quality images of internal organs that are close to the gastrointestinal tract \cite{ang2018diagnostic}. A small US probe is attached to an endoscope, which is inserted through the mouth or rectum to visualize organs such as the pancreas, liver, and lungs. EUS is commonly used to assess digestive diseases, guide biopsies, and evaluate tumors, particularly in the pancreas and esophagus.
    
    \item \textbf{Doppler ultrasound} is a specialized technique that measures the movement of blood through vessels by detecting changes in the frequency of the reflected sound waves, known as the Doppler effect \cite{routh1996doppler}. It provides critical information about blood flow, including velocity and direction, helping detect conditions such as blockages, clots, or reduced blood flow due to narrowing of the arteries.

    \item \textbf{Transvaginal ultrasound} is a type of pelvic US where a probe is inserted into the female genitalia to obtain detailed images of the uterus, ovaries, cervix, and surrounding structures \cite{fossum1988early}. This method offers higher resolution than abdominal US due to the closer proximity of the probe to the pelvic organs. It is commonly used for early pregnancy evaluations, diagnosing ovarian cysts, and assessing abnormal bleeding or pelvic pain.
    
\end{itemize}

\begin{figure}
    \centering
    \includegraphics[width=1\linewidth]{us.jpg}
    \caption{US image of the fetus at 12 weeks of pregnancy in a sagittal scan}
    \label{fig:us}
\end{figure}


\section{U-Net and its variants}\label{sec4}

U-Net is a convolutional neural network (CNN) primarily designed for biomedical image segmentation \cite{ronneberger2015u}. The architecture was proposed by Ronneberger et al. in 2015 and is widely used for pixel-level tasks due to its ability to capture both global context and fine details. Its symmetric encoder-decoder structure allows it to effectively model complex features while preserving spatial information.

\subsection{U-Net Architecture}

\begin{figure}
    \centering
    \includegraphics[width=1.0\linewidth]{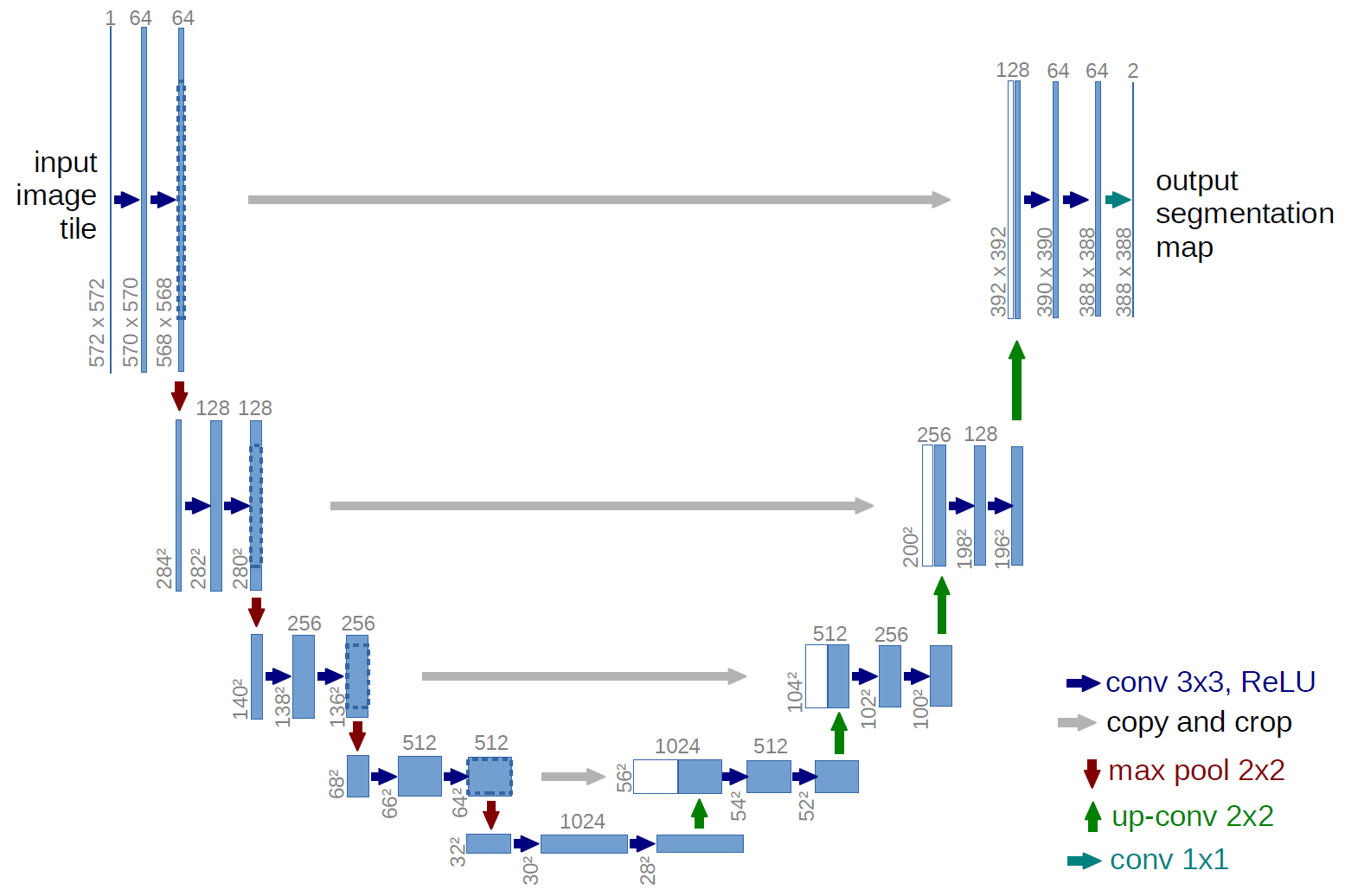}
    \caption[U-Net Architecture]{\textls[-15]{U-Net Architecture \cite{ronneberger2015u}. The U-Net begins with a 572x572 input image. The encoder path applies two 3x3 convolutions with ReLU activations (blue arrows) in each block, increasing feature channels from 64 to 512 while reducing spatial dimensions via 2x2 max-pooling (red arrows). The bottleneck layer has 1024 channels and a 28x28 size. In the decoder, 2x2 up-convolutions (green arrows) double the spatial dimensions and halve the channels, combining with corresponding encoder features through skip connections (copy and crop-grey arrows) for better localization. A final 1x1 convolution (teal arrow) outputs a 388x388 segmentation map with two channels.}}
    \label{fig:unet}
\end{figure}

The U-Net architecture consists of two main parts: the \textit{encoder} (contracting path) and the \textit{decoder} (expanding path)  as shown in Fig. \ref{fig:unet}.

\subsubsection{Encoder (Contracting Path)}

The encoder consists of series of convolutional layers followed by max-pooling layers. Each convolutional layer applies a convolution operation with filters of size $k \times k$, using optional padding to preserve the spatial dimensions. The output of each convolutional block is subsequently passed through a non-linear activation function, rectified linear unit (ReLU) or Gaussian Error Linear Unit (GeLU). Mathematically, the output of a convolutional layer can be expressed as:
\[
Y = \sigma(W \ast X + b)
\]
where $X \in \mathbb{R}^{H \times W \times C}$ denote the input image, where $H$, $W$, and $C$ represent the height, width, and number of channels, respectively. $W$ represents the convolutional kernel, $b$ is the bias, $\ast$ denotes the convolution operation, and $\sigma$ is ReLU/GeLU. After each convolution, a $m \times m$ max-pooling operation is applied to reduce the spatial resolution by a factor of $m$.

\subsubsection{Bottleneck}

The bottleneck, or bridge, connects the encoder and decoder. It consists of $cl$, $k \times k$ convolutions, followed by a ReLU/GeLU activation function. This layer captures the deepest level of feature representations with the smallest spatial dimension.

\subsubsection{Decoder (Expanding Path)}

The decoder is structurally symmetric to the encoder and comprises upsampling operations followed by convolutional layers. The upsampling operation doubles the spatial resolution, achieved through either transposed convolution (also known as deconvolution) or interpolation techniques.  This can be represented as:
\[
Z = W^T \ast Y
\]
where $W^T$ is the transposed convolution kernel, and $Y$ is the input from the previous layer. After upsampling, the corresponding feature map from the encoder is concatenated to the decoder feature map to preserve spatial information. This is known as a skip connection and is crucial for retaining fine details.

Each concatenated feature map is then passed through $cl$, $k \times k$ convolutions followed by activation function, progressively reconstructing the spatial resolution while refining the feature map.

\subsubsection{Output Layer}

The final layer of the U-Net architecture is a $1 \times 1$ convolution which reduces the number of channels to the desired number of output classes. For segmentation tasks, the softmax function is often applied to obtain a probability distribution over the classes for each pixel:
\[
P(c_i | X) = \frac{\exp(s_i)}{\sum_{j=1}^{C}\exp(s_j)}
\]
where $s_i$ is the score for class $i$, and $C$ is the total number of classes.

\subsubsection{Loss Function}

For binary segmentation tasks, U-Net uses the binary cross-entropy (BCE) loss:
\[
\mathcal{L}_{\text{BCE}} = -\frac{1}{N}\sum_{i=1}^{N} \left[ y_i \log(\hat{y}_i) + (1 - y_i) \log(1 - \hat{y}_i) \right]
\]
where $y_i$ is the true label, $\hat{y}_i$ is the predicted probability, and $N$ is the total number of pixels. For multi-class segmentation, CE loss is generalized as:
\[
\mathcal{L}_{\text{CE}} = -\sum_{i=1}^{N} \sum_{c=1}^{C} y_{i,c} \log(\hat{y}_{i,c})
\]
where $y_{i,c}$ is the one-hot encoded label for class $c$ and $\hat{y}_{i,c}$ is the predicted probability for class $c$.

\subsection{U-Net++}

U-Net++ is an advanced variant of the U-Net architecture proposed by Zhou et al. in 2018 \cite{zhou2018unet++}. It enhances semantic segmentation performance by redesigning skip connections and introducing dense convolutional blocks between the encoder and decoder. U-Net++ achieves higher accuracy through \textit{nested convolutional pathways} that facilitate feature refinement and multi-scale feature fusion.

\subsubsection{U-Net++ Architecture}

\begin{figure}
    \includegraphics[width=1.0\linewidth]{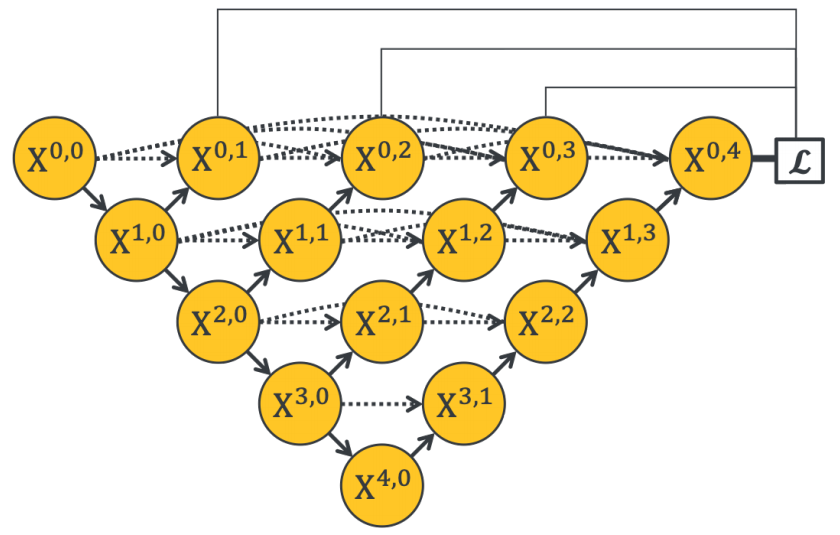}
    \caption{\textls[-15]{U-Net++ Architecture \cite{zhou2018unet++}. This figure illustrates the nested convolutional pathways in U-Net++. Each node \( X^{i,j} \) represents the output of a convolutional block at the \( i \)-th encoder/decoder level and \( j \)-th convolutional layer within the nested pathway. Solid arrows indicate primary connections between encoder and decoder layers, while dotted arrows represent nested skip connections, enabling multi-scale feature fusion and progressive feature refinement. The final output \( X^{0,4} \) is computed after aggregating features across multiple depths, and the symbol \( \mathcal{L} \) denotes the loss function, which is used to optimize the network for improved segmentation accuracy.
}}
    \label{fig:unet++}
\end{figure}

The architecture of U-Net++ retains the fundamental encoder-decoder structure of the original U-Net and introduces significant modifications in the skip connections. Specifically, it employs nested convolutional pathways, which consist of a series of convolutional blocks connecting encoder and decoder layers at various depths, as illustrated in Fig. \ref{fig:unet++}.

\subsubsection{Nested Convolutional Pathways}

In U-Net++, skip connections are redefined to include convolutional blocks that progressively refine features before they are merged with decoder features. Let $X^{i,j}$ denote the feature map at the $i$-th decoder stage and $j$-th convolutional layer within the nested pathway. The feature maps are computed recursively as:

\[
X^{i,j} = 
\begin{cases}
    f\left( X^{i-1,j}, \ \mathrm{Up}\left( X^{i,j-1} \right) \right), & \text{if } j > 0 \\
    f\left( X_{\text{enc}}^{i} \right), & \text{if } j = 0
\end{cases}
\]

where:
\begin{itemize}
    \item $X_{\text{enc}}^{i}$ is the output feature map from the $i$-th encoder layer.
    \item $f(\cdot)$ represents a convolutional operation (e.g., convolution followed by batch normalization and activation).
    \item $\mathrm{Up}(\cdot)$ denotes an upsampling operation to match the spatial dimensions.
\end{itemize}

The nested pathways allow the network to aggregate features from different semantic scales, effectively bridging the semantic gap between encoder and decoder features. This design enhances the representational capacity of the network by enabling deeper supervision and more precise feature alignment.

Each decoder node $X^{i,j}$ is connected not only to its corresponding encoder feature map $X_{\text{enc}}^{i}$ but also to all preceding decoder nodes $X^{k,j-1}$ where $k < i$. This dense connectivity can be visualized as a full convolutional block between encoder and decoder stages, promoting extensive feature reuse and refinement.

The overall output of the network is obtained from the deepest decoder layer after applying a final convolutional layer to map the feature maps to the desired number of segmentation classes.

\subsection{U-Net 3+}

U-Net 3+ is an enhanced version of the U-Net \cite{ronneberger2015u} and U-Net++ \cite{zhou2018unet++} architectures, introduced by Huang et al. in 2020 to enhance multi-scale feature fusion and segmentation accuracy, particularly for pixel-wise prediction tasks \cite{huang2020unet}. U-Net 3+ introduces two main innovations: \textit{full-scale skip connections} and \textit{deep supervision}. These modifications enable the integration of information across all encoder and decoder layers, enhancing the network's ability to capture both high-level semantic information and low-level spatial details.

\subsubsection{U-Net 3+ Architecture}

\begin{figure}
    \includegraphics[width=1.0\linewidth]{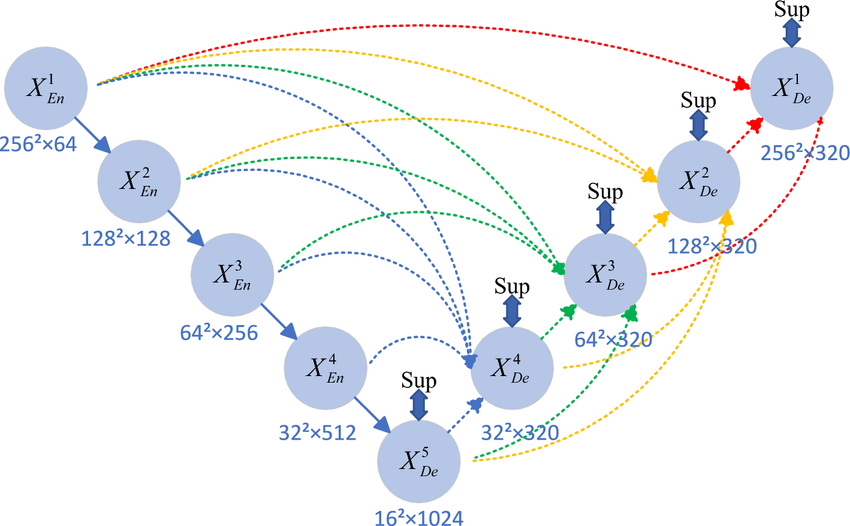}
    \caption{\textls[-15]{U-Net 3+ Architecture \cite{huang2020unet}. Given figure shows the use of full-scale skip connections and deep supervision in U-Net 3+. Each encoder feature map \( X^{i}_{En} \) and decoder feature map \( X^{j}_{De} \) is labeled with spatial dimensions and number of channels (e.g., \( 256^2 \times 64 \) for \( X^{1}_{En} \) and \( 256^2 \times 320 \) for \( X^{1}_{De} \)). The notation \( 256^2 \times 64 \) indicates a spatial resolution of \( 256 \times 256 \) with 64 channels. Full-scale skip connections (color-coded dotted lines) link encoder feature maps at different resolutions with all corresponding decoder levels, allowing multi-scale feature fusion by combining low-level spatial details with high-level semantic information. Deep supervision layers ('Sup') are applied to each decoder level to provide auxiliary output supervision, enhancing feature learning at multiple scales and improving segmentation accuracy.}}
    \label{fig:unet3+}
\end{figure}

The U-Net 3+ architecture maintains the basic encoder-decoder structure of U-Net and redefines the skip connections. In U-Net 3+, each decoder level aggregates feature maps from all encoder levels via \textit{full-scale skip connections}. This design facilitates the fusion of features from multiple resolutions, as illustrated in Fig. \ref{fig:unet3+}.

\subsubsection{Full-Scale Skip Connections}

U-Net 3+ employs full-scale skip connections to aggregate feature maps from all encoder layers into each decoder layer. Let \( Z_{i,j} \) represent the feature map at the \( i \)-th level of the decoder after fusion with the encoder outputs for level \( j \). The full-scale skip connections are defined by:

\[
Z_{i,j} = \text{concat}(E_0, E_1, \dots, E_N, D_{i-1,j})
\]

where:
\begin{itemize}
    \item \( E_k \) is the feature map from the \( k \)-th encoder level, \( k \in \{0, 1, \dots, N\} \),
    \item \( D_{i-1,j} \) is the upsampled feature map from the previous decoder level \( i-1 \), and
    \item \text{concat} refers to the concatenation operation across all encoder features and the corresponding decoder feature map.
\end{itemize}

This concatenation allows each decoder layer to access and integrate information from multiple resolution levels, enhancing the model's ability to capture diverse features and improving segmentation precision.

\subsubsection{Deep Supervision}

In addition to full-scale skip-connections, U-Net 3+ incorporates a \textit{deep supervision} mechanism that applies auxiliary output layers to intermediate decoder levels. For each decoder level \( D_i \), an auxiliary output \( X_{\text{output}}^{(i)} \) is generated, and an associated loss is calculated. The total loss function, \( \mathcal{L}_{\text{total}} \), combines the individual losses from all decoder levels, encouraging learning across multiple scales. This is formulated as:

\[
\mathcal{L}_{\text{total}} = \sum_{i=1}^{M} \lambda_{i} \, \mathcal{L}\left(X_{\text{output}}^{(i)}, Y\right)
\]

where:
\begin{itemize}
    \item \( Y \) is the ground truth segmentation map,
    \item \( \mathcal{L} \) denotes the segmentation loss function, typically pixel-wise cross-entropy,
    \item \( X_{\text{output}}^{(i)} \) is the predicted segmentation map at the \( i \)-th decoder level, and
    \item \( \lambda_{i} \) are weights for each decoder level’s contribution to the total loss.
\end{itemize}

The deep supervision mechanism ensures that each decoder layer is optimized individually, enhancing multi-scale feature learning and improving the final segmentation output. The model’s final segmentation map is produced by combining these supervised outputs from each level.

\begin{table*}[H]
    \centering
    \caption{Summary of U-Net Integrated with X-ray}
    \label{tab:xray}
    \resizebox{\textwidth}{!}{ 
        \begin{tabular}{@{}>{\raggedright}p{3cm} >{\raggedright}p{3.5cm} >{\raggedright}p{4cm} >{\raggedright}p{4cm} >{\raggedright\arraybackslash}p{4cm}@{}}
            \toprule
            \textbf{Study} & \textbf{Modality} & \textbf{Focus Area} & \textbf{Methodology} & \textbf{Performance Metrics} \\ 
            \midrule
            Deng et al. (2024) \cite{deng2024efficient} & X-ray (Anterior-Posterior and Lateral) & Vertebrae instance segmentation for spinal disorder diagnosis & Enhanced U-Net architecture using ConvNeXt as encoder, Informational feature enhancement (IFE) module for texture and edge enhancement, attention in bottleneck, and Residual Network (ResNet) blocks in decoder. & Accuracy: 88.0\%, Dice Similarity Coefficient (DSC): 90.6\%, Mean intersection over Union (IoU): 79.3\% \\ 
            
            Haennah et al. (2024) \cite{haennahcombined} & Chest X-ray (CXR) & Early diagnosis of Severe Acute Respiratory Syndrome Coronavirus 2 (SARS-CoV-2) & Classification with Fused U-Net Convolutional Neural Network (FUCNN) optimized by Chaotic System-based Moth Flame Optimization (CSMFO) & Accuracy: 98.5\%, Sensitivity: 98.6\%, Specificity: 98.9\%, Precision: 98.9\% \\ 
            
            Sharma et al. (2024) \cite{sharma2024deep} & Chest X-ray (CXR) & Tuberculosis Detection & U-Net for lung segmentation, Xception for classification, with gradient-weighted class activation mapping (Grad-CAM) for visualization & Segmentation: Accuracy: 96.5\%, Jaccard Index: 90.4\%, Dice Coefficient Index (DCI): 94.8\%, Classification: Accuracy: 99.3\%, Precision: 99.3\%, Recall: 99.3\%, F1-score: 99.3\% \\

            Ying et al. (2024) \cite{ying2024performance} & Clinical X-ray & Dental Caries Detection & Comparison of object detection: You Only Look Once version 5 (YOLOv5), Detection Transformer (DETR), and segmentation networks: U-Net, and transformer-based U-Net & F1-score: YOLOv5 (87.0\%), DETR (82.0\%); U-Net (80.0\%); transformer-based U-Net (86.0\%) \\

            Budagam et al. (2024) \cite{budagam2024instance} & Dental X-ray (Panoramic) & Teeth segmentation and recognition & U-Net and YOLO version 8 (BB-UNet) & mean average precision (mAP): 72.9\%, Precision: 94.3\%, Recall: 92.3\%, DCI: 84.0\% \\ 

            Lyu et al. (2023) \cite{lyu2023mwg} & Chest X-ray & Lung and heart segmentation & Multiple tasking Wasserstein generative adversarial network U-Net & Dice Similarity: 95.3\%, Precision: 96.4\%, F1-score: 95.9\%, IoU (Lung): 81.4\%, IoU (Heart): 74.6\%, DCI (Lung): 85.2\%, DCI (Heart): 71.2\% \\

            Wu et al. (2021) \cite{wu2021ulnet}& Chest X-ray (CXR) & COVID-19 Detection & Modified U-Net-based CNN model for binary classification (COVID-19 vs. Normal) and multiclass classification (COVID-19 vs. Normal vs. Viral Pneumonia) & Binary classification: Accuracy: 99.5\%; Multiclass classification: Accuracy: 95.4\% \\

            Mosquera-Berrazueta et al. (2023) \cite{10178991} & Chest X-ray (CXR) & Tuberculosis lesion detection and segmentation & an optimized U-net variant, using ten-fold stratified cross-validation & DCI: 92.0\%, IoU: 86.0\% \\

            Agarwal et al. (2023) \cite{agrawal2023rese} & Chest X-ray (CXR) and CT-scan & Lung segmentation & Proposed a UNet-based model incorporating residual learning and attention mechanisms; & average DCI: 96.4\%; Average Jaccard Index (JI): 93.1\% \\

            Kholiavchenko et al. (2020) \cite{kholiavchenko2020contour} & Chest X-ray (CXR) & Organ segmentation (lung fields, heart, clavicles) & Augmented state-of-the-art CNNs (UNet, LinkNet with ResNeXt, Tiramisu with DenseNet) with organ contour information; & JI: 97.1\% (lung fields), 93.3\% (heart), 90.3\% (clavicles) \\

            \bottomrule
        \end{tabular}
    } 
\end{table*}

\section{U-Net integration across various medical imaging modalities}\label{sec5}

\begin{figure}
    \includegraphics[width=1\linewidth]{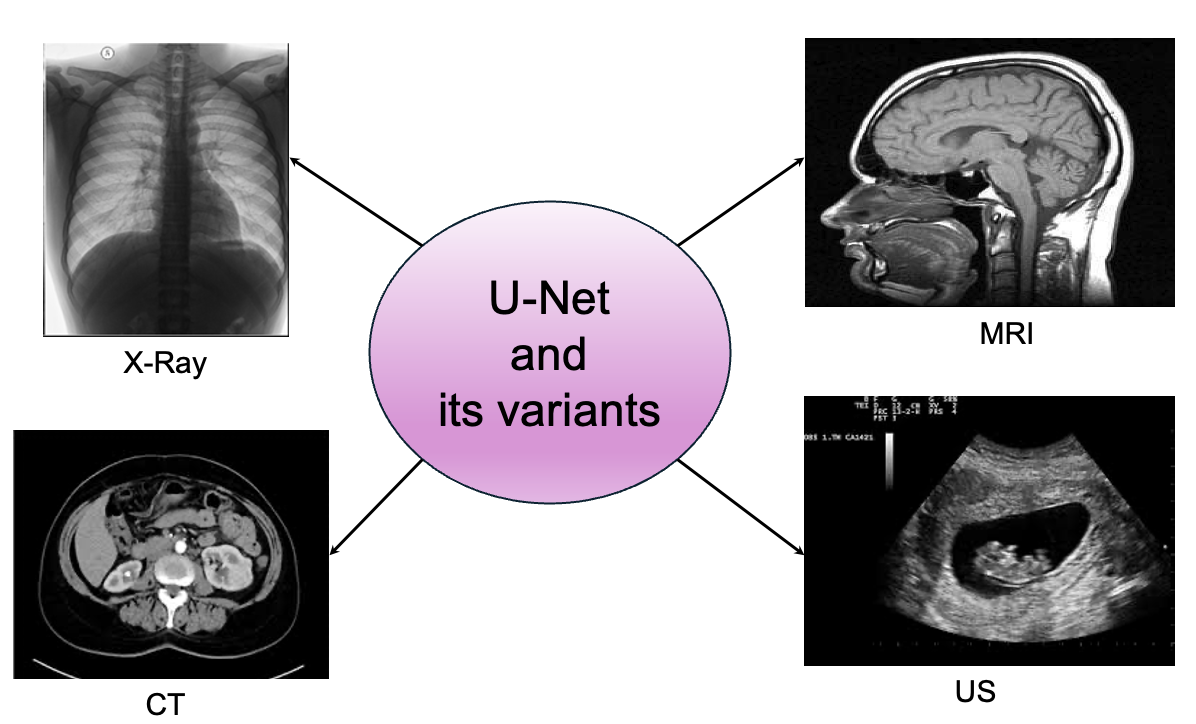}
    \caption{\textls[-15]{U-Net integration across various medical imaging modalities}}
    \label{fig:unetimage}
\end{figure}

The integration of U-Net architectures in healthcare has transformed the way medical images are analyzed across various modalities, see Fig. \ref{fig:unetimage}. Designed specifically for tasks like segmentation, U-Net has proven highly adaptable, consistently delivering precise results in identifying and separating different structures within complex images. This section explores the application of U-Net in various imaging modalities, highlighting its effectiveness in enhancing diagnostic accuracy and supporting clinical decision-making. 

\subsection{U-Net Integration with X-ray}
X-ray imaging, known for its accessibility and efficiency, gains enhanced diagnostic power through U-Net integration. Table \ref{tab:xray} presents a summary of recent studies, applying U-Net in X-ray analysis, outlining focus areas, methodologies, and performance metrics.

\renewcommand{\arraystretch}{1.5} 

\subsection{U-Net Integration with MRI}

MRI imaging gains improved segmentation accuracy with U-Net integration. Table \ref{tab:mri} summarizes recent studies applying U-Net to MRI analysis.

\renewcommand{\arraystretch}{1.5} 

\begin{table*}[htbp] 
    \centering
    \caption{Summary of U-Net Integrated with MRI}
    \resizebox{\textwidth}{!}{ 
    \begin{tabular}{@{}>{\raggedright}p{3cm} >{\raggedright}p{3.5cm} >{\raggedright}p{4cm} >{\raggedright}p{4cm} >{\raggedright\arraybackslash}p{4cm}@{}}
        \toprule
        \textbf{Study} & \textbf{Modality} & \textbf{Focus Area} & \textbf{Methodology} & \textbf{Performance Metrics} \\ 
        \midrule
        
        Yeboah et al. (2024) \cite{yeboah2024enhancing} & Brain MRI & Lesion segmentation & Transfer learning with U-Net and Feature Pyramid Networks (FPN) architectures & DSC: 98.1\% \\
        
        Wang et al. (2024) \cite{wang2024supraspinatus} & Supraspinatus (Shoulder MRI) & Muscle segmentation & Attention-dense atrous spatial pyramid pooling U-Net with ResNet version 34 as an encoder & Precision: 99.2\%; IoU: 83.4\%; DICE: 91.0\% \\
        
        Hossain et al. (2024) \cite{hossain2024swin} & Brain MRI & Motion artifact correction & U-Net with Swin Transformer: Multiscale contextual feature-extraction with dual upsampling to improve spatial resolution in the decoder. & Mean SSIM: 91.7\% \\

        Das et al. (2024) \cite{das2024attention} & Cardiac MRI (C-MRI) & Cardiac ventricle segmentation & Attention-UNet models with and without pretrained backbones (ResNet50, DenseNet121) & DCI: 98.9\% (Attention-UNet), 97.2\% (ResNet50), 98.0\% (DenseNet121); IoU: 97.8\%, 94.6\%, 96.1\% respectively.\\

        Borra et al. (2024) \cite{borra2024automatic} & Brain MRI & Brain tumor detection and classification & Edge detection with U-Net for segmentation; Support Vector Machine (SVM) for classification & Accuracy: 99.7\%; Specificity: 99.7\%; Precision: 98.8\%; Sensitivity: 97.4\% \\

        Wang et al. (2023) \cite{wang2023deep} & Upper abdomen MRI & Hepatocellular carcinoma segmentation & U-Net++ & DSC (Liver): 92.0\%; DSC (Tumors): 68.7\%; \\
        
        Wang et al. (2023) \cite{wang2023mlkca} & Sagittal T2-weighted MRI & Spinal disease segmentation & Multiscale large-kernel convolution Attention U-Net (MLKCA-U-Net) & IoU: 83.0\%; DSC: 90.2\% \\
        
        Dolz et al. (2018) \cite{dolz2018ivd} & 3D multi-modal lower spine MRI & Intervertebral disc localization and segmentation & IVD-Net: Extends U-Net with multi-modal MRI data and densely connected paths inspired by HyperDenseNet & mean DSC: 91.6\%\\

        Jia et al. (2022) \cite{jia2021bitr}& 3D brain MRI (T1-weighted (T1), post-contrast T1-weighted (T1c), T2-weighted (T2), and T2 Fluid Attenuated Inversion Recovery (T2-FLAIR))&  Brain Tumor Segmentation & a CNN-Transformer based U-Net to leverage long-range feature extraction & Median Dice scores: 93.4\% (whole tumor), 93.0\% (tumor core), 88.9\% (enhancing tumor) \\

        Thomas et al. (2022) \cite{thomas2020multi} & 3D FLAIR weighted MRI & Focal Cortical Dysplasia (FCD) Segmentation & Multi-Res-Attention UNet with a hybrid skip connection-based fully convlutional network (FCN) architecture & FCD detection rate (Recall): 92.0\% \\

        \bottomrule
    \end{tabular}
    } 
    \label{tab:mri}
\end{table*}

\subsection{U-Net Integration with CT scan}
U-Net integration with CT scans improves segmentation accuracy, supporting more detailed and reliable diagnostic insights. This combination is particularly effective for detecting and delineating complex structures, such as tumors, organs, and tissues, where precise segmentation is critical for treatment planning and assessment. Table \ref{tab:ct} summarizes recent work showcasing the advancements in U-Net and CT scan integration across various clinical applications.

\renewcommand{\arraystretch}{1.5} 

\begin{table*}[h!] 
    \centering
    \caption{Summary of U-Net Integrated with CT Scan}
    \resizebox{\textwidth}{!}{ 
    \begin{tabular}{@{}>{\raggedright}p{3cm} >{\raggedright}p{3.5cm} >{\raggedright}p{4cm} >{\raggedright}p{4cm} >{\raggedright\arraybackslash}p{4cm}@{}}
        \toprule
        \textbf{Study} & \textbf{Modality} & \textbf{Focus Area} & \textbf{Methodology} & \textbf{Performance Metrics} \\ 
        \midrule

        Morani et al. (2024) \cite{morani2024detecting} & CT scans & COVID-19 Diagnosis & U-Net based segmentation module (with optional slice removal), lung extraction, and classification module & Average Dice Score: 97.0\% \\

        Chauhan et al. (2024) \cite{chauhan2024unet} & CT (low dose) scans & CT Image Denoising & U-Net based architecture combined with ConvNeXt features (ResNextify and inverted bottleneck), enhanced denoising capabilities by addressing residual noise and preserving structural details through a generator & SSIM: 97.6\% \\

        Neha et al. (2024) \cite{neha2024multi} & CT scans & Renal Tumor Segmentation & U-Net model with residual connections, multi-layer feature fusion (MFF), and cross-channel attention (CCA) within encoder blocks & DSC: 97.0\% and JI: 95.0\% for kidney segmentation; DSC: 96.0\% and JI: 91.0\% for renal tumor segmentation \\

        Ou et al. (2024) \cite{ou2024restransunet} & CT scans & Liver Segmentation & ResTransUNet: A model combining U-Net and Transformer architectures with a feature enhancement unit to capture global context and spatial relationships, combining Residual (Squeeze and Excitation) SE-block, Swin Transformer, ASPP, and Feature Enhancement Unit (FEU) & Achieved DCI: 95.4\% \\

        Çelebi et al. (2024) \cite{ccelebi2024maxillary} & CT scans - Cone Beam (CB) & Maxillary Sinus Detection & Res-Swin-UNet: A U-Net architecture combining ResNet and Swin Transformer & Achieved F1-score: 91.7\%, Accuracy: 99.0\%, IoU: 84.7\% \\

        Kamanli et al. (2024) \cite{kamanli2024hyperparameter} & CT scans (contrast-agent-free) & Ischemic and Hemorrhagic Stroke Detection & Enhanced U-Net model integrated with Cross Patch Attention Module (CPAM) & Classification Accuracy: 95.0\%, IoU: 88.0\% \\

        Lei et al. (2024) \cite{lei2024adipose} & Chest CT scans & Lung Adipose Tissue Detection & ConvBiGRU: A model for lung slice localization and a multi-module U-Net-based model for segmenting subcutaneous (SAT) and visceral adipose tissue (VAT) & Achieved DSC: 92.0\% (SAT) and 82.7\% (VAT); F1 Scores: 82.2\% (SAT) and 78.8\% (VAT) \\

        Neha et al. (2023) \cite{10406429} & CT scans & Kidney Tumor Segmentation & Dense SIFT-integrated U-Net-based network: Utilized DenseSIFT images as input in a U-Net encoder-decoder architecture & Mean IoU: 91.9\% \\

        Gillot et al. (2022) \cite{gillot2022automatic} & CBCT scans & Full-Face Segmentation for Clinical Decision Support & UNETR: U-Net with Transformers & Dice Score: 96.2\% \\

        Yousefi et al. (2021) \cite{yousefi2021esophageal} & CT scans & Esophageal Tumor Segmentation & DDAUnet: Dilated Dense Attention U-Net with spatial and channel attention gates, leveraging dilated convolutional layers to expand the receptive field and optimize memory use & DSC: 79.0\% \\

        \bottomrule
    \end{tabular}
    } 
    \label{tab:ct}
\end{table*}

\subsection{U-Net Integration with Ultrasound}

Ultrasound imaging, valued for its real-time and non-invasive capabilities, benefits from enhanced segmentation accuracy through U-Net integration as shown in the following table \ref{tab:us}.

\renewcommand{\arraystretch}{1.5} 

\begin{table*}[h!]
    \centering
    \caption{Summary of U-Net Integrated with Ultrasound}
    \resizebox{\textwidth}{!}{ 
    \begin{tabular}{@{}>{\raggedright}p{3.5cm} >{\raggedright}p{3.5cm} >{\raggedright}p{4cm} >{\raggedright}p{4cm} >{\raggedright\arraybackslash}p{4cm}@{}}
        \toprule
        \textbf{Study} & \textbf{Modality} & \textbf{Focus Area} & \textbf{Methodology} & \textbf{Performance Metrics} \\ 
        \midrule

        Malekmohammadi et al. (2024) \cite{malekmohammadi2024mass} & Breast Ultrasound Imaging & Breast Cancer Detection & Bi-ConvLSTM U-Net with Convolutional Block Attention Module (CBAM) for automatic segmentation & DSI: 85.8\% \\

        Inan et al. (2024) \cite{inan2024multi} & Ultrasonography Images & Thyroid Nodule Segmentation and Classification (AUS/FLUS, benign follicular, papillary follicular) & Hybrid AI-based system integrating ResUNet and ResUNet++ for segmentation and classifiers (VGG-16, DenseNet121, ResNet-50, Inception ResNet-v2) for nodule classification & DCI: 92.4\%, mean IoU: 89.7\% (ResUNet++), classification accuracy: 96.6\% (ResNet-50), 97.0\% (AUS/FLUS) \\

        Luo et al. (2024) \cite{luo2024rpa} & Ultrasound Images & Arteriovenous Fistula (AVF) Segmentation & RPA-UNet: Residual Pyramidal Attention U-Net with attention mechanisms & IoU: 91.4\%, Recall: 97.2\%, Dice: 95.3\%, Precision: 93.7\% \\

        Jiang et al. (2024) \cite{jiang2024microsegnet} & Micro-Ultrasound Imaging & Prostate Segmentation & MicroSegNet: Multiscale annotation-guided transformer U-Net & DCI: 93.9\% \\

        Sarkar et al. (2024) \cite{sarkar2024dc} & Ovarian Ultrasound Images & Automatic Ovarian Follicle Segmentation & DC-UNet: Double Contraction U-Net with two contracting paths & Accuracy: 97.8\%, Precision: 97.5\%, Recall: 94.3\%, F1 Score: 95.9\%, DSC: 76.0\%, JI: 59.0\% \\

        Chang et al. (2024) \cite{chang2024wrist} & Wrist Joint Ultrasound Images & Rheumatoid Arthritis Detection & SEAT-UNet: U-Net with self-attention mechanism for synovial hypertrophy and effusion detection & Sensitivity: 100\%, DCI: 84\% (synovial hypertrophy); Sensitivity: 86\%, DCI: 84\% (effusion) \\

        Zhang et al. (2024) \cite{zheng2024radiological} & Knee Joint Ultrasound Images & Knee Osteoarthritis Diagnosis & Improved Unet3+ with attention mechanisms and ASPP & Dice accuracy: 78.7\%, average area accuracy: 91.1\%, average distance accuracy: 91.1\% \\

        Hao et al. (2024) \cite{hao2024new} & Ultrasound Images & Carotid Artery Plaque Segmentation & RCSU-Net: Enhanced U-Net with residual convolution, CBAM, and multi-scale supervision & DSC: 81.9\%, IoU: 70.3\% \\

        Ejiyi et al. (2024) \cite{ejiyi2024attention} & Breast Ultrasound Images (BUSI) and Retinal Fundus Images (RFI) & Computer-Aided Diagnosis (CAD) & ADU-Net: Attention-Enriched Deeper U-Net with global context and progressive context refinement modules & mIoU: 76.5\% (BUSI), 59.2\% (RFI); F1 scores: 62.1\% (BUSI), 71.7\% (RFI); accuracy: 94.9\% (BUSI), 96.0\% (RFI) \\

        Li et al. (2019) \cite{li2019cr} & Transvaginal Ultrasound & Ovarian and Follicle Segmentation & CR-U-Net: Spatial recurrent neural network-integrated with U-Net architecture & DSC: 91.2\% (ovary), 85.8\% (follicles) \\

        \bottomrule
    \end{tabular}
    } 
    \label{tab:us}
\end{table*}

\section{Limitations} \label{sec6}

U-Net and its variants are widely used in medical image segmentation. This section discusses the limitations of U-Net, U-Net++, and U-Net3+, as well as those of X-ray, MRI, CT, and Ultrasound (US) imaging modalities.

\subsection{U-Net}
U-Net's fixed-size convolutional kernels and pooling layers limit its receptive field, reducing the ability to capture long-range dependencies essential for segmenting large or complex structures. Downsampling leads to loss of fine spatial details, and skip-connections do not fully recover this information, affecting segmentation of small or intricate features.

\subsection{U-Net++}
U-Net++ enhances feature propagation with nested and dense skip connections, reducing the semantic gap between encoder and decoder features. However, this increases model complexity and the number of parameters significantly, leading to longer training times and greater computational resource requirements, which may be impractical for real-world applications. The higher parameter count elevates the risk of overfitting, especially on small datasets, degrading generalization performance on unseen data.

\subsection{U-Net3+}
U-Net3+ incorporates full-scale skip connections and deep supervision to integrate multi-scale features effectively. While this captures both high-level semantic information and low-level spatial details, it significantly increases architectural complexity and computational burden. The intricate architecture with multiple pathways and supervisory signals complicates the interpretability of the model. This makes it harder to analyze the contribution of each component to the final output, which is important for clinical validation and trust in medical applications.

\subsection{X-Ray}
X-ray imaging has poor soft-tissue contrast as it relies on differences in tissue density. This limitation makes it difficult to distinguish between soft tissues with similar densities, complicating accurate segmentation of organs or lesions. X-ray images are two-dimensional projections of three-dimensional structures, resulting in overlapping anatomical features and loss of depth information. This projection effect obscures critical details and complicates the segmentation task. Noise and artifacts, such as scatter radiation and motion blur, further degrade image quality and hinder segmentation algorithms.

\subsection{MRI}
MRI provides excellent soft-tissue contrast and multi-planar imaging capabilities. However, it is susceptible to artifacts such as magnetic susceptibility near metallic implants, causing signal voids or distortions. High operational costs and specialized equipment limit accessibility, resulting in smaller datasets for training models. Variations in MRI signals between scanners and imaging protocols lead to domain shift problems, affecting model generalization.

\subsection{CT}
CT imaging involves exposure to ionizing radiation, raising concerns about cumulative doses, especially for children and patients requiring multiple scans. Artifacts like beam hardening and motion introduce distortions that obscure anatomical details and complicate segmentation. Limited soft-tissue contrast compared to MRI makes differentiating soft tissues challenging without contrast agents.

\subsection{US}
Ultrasound imaging is safe, portable, and inexpensive but image quality is highly operator-dependent, leading to inconsistencies that hinder model generalization. Artifacts such as speckle noise, shadowing, and reverberations degrade image quality and obscure structures. Limited field of view and lower spatial resolution hinder visualization of large or deep structures.

\section{Discussion and Future Directions} \label{sec7}

Overcoming the limitations of U-Net variants and medical imaging modalities requires a multifaceted approach that emphasizes efficiency, generalization, and adaptability. While U-Net has demonstrated outstanding performance in medical image segmentation, several strategies can be implemented to address current challenges and enhance its capabilities for real-world clinical applications.

To create efficient models that can be deployed in resource-constrained environments, techniques such as model pruning, quantization, knowledge distillation, and the integration of attention mechanisms can be leveraged. These methods not only reduce computational overhead but also help capture critical features and long-range dependencies. Additionally, employing graph neural networks (GNNs) or dilated convolutions can improve feature representation and maintain computational efficiency, allowing U-Net models to better capture complex structures in medical images.

Improving image quality and standardizing imaging protocols across different modalities is another critical challenge. Variations in image quality and protocols across X-ray, CT, MRI, and ultrasound modalities can introduce inconsistencies that complicate segmentation tasks. Implementing advanced artifact correction algorithms, motion compensation techniques, and harmonizing imaging protocols across different modalities can significantly reduce these sources of variability, ultimately improving model generalization and enabling U-Net models to work more effectively across diverse imaging types.

One of the most pressing issues in medical imaging is the scarcity of large, labeled datasets. To address this, Generative AI techniques such as Generative Adversarial Networks (GANs) and Variational Autoencoders (VAEs) offer promising solutions. These methods can generate synthetic medical images that augment existing datasets, helping to increase data diversity and reduce overfitting. By enhancing the availability of training data, these techniques improve model robustness and generalization, making U-Net models more effective in clinical practice.

Furthermore, integrating domain-specific knowledge, such as radiomics, pathological, and serological information, is essential for improving the clinical relevance of U-Net-based models. By incorporating known anatomical structures and contextual information into the model’s training, U-Net can be adapted to produce more clinically meaningful segmentation results. In addition, combining multimodal data from different imaging techniques or clinical textual data can provide richer, more comprehensive information, leading to more accurate and reliable model predictions.

Adapting Transformer-based architectures to multimodal learning, where imaging data is combined with clinical textual information, is another area of significant potential. These architectures, which were originally developed for natural language processing, have been successfully applied to vision tasks and can model global dependencies within medical image data. By capturing contextual details from clinical notes or radiology reports alongside imaging data, these models can enhance segmentation accuracy and improve the clinical decision-making process.

Finally, the integration of explainable AI (XAI) techniques is crucial for the widespread acceptance of U-Net models in clinical settings. Clinicians require transparent and interpretable models to trust and effectively incorporate AI-driven results into their workflows. By utilizing XAI approaches, such as saliency maps or attention mechanisms, model decisions can be better elucidated, fostering greater trust among clinicians and improving the integration of these models into clinical environments.

\section{Conclusion} \label{sec8}
In this paper, we reviewed the most widely used medical imaging modalities—X-ray, MRI, CT, and Ultrasound—and explored the application of U-Net and its variants for medical image segmentation. We provided an in-depth analysis of the architectures of these models and examined recent studies that have integrated U-Net with these imaging modalities. By discussing the limitations of current approaches, we highlighted the key challenges faced in the field, including issues related to data scarcity, image quality variability, and model generalization across modalities.

Additionally, we proposed effective strategies to address these challenges, focusing on enhancing model efficiency, improving data diversity, and incorporating domain-specific knowledge. Our review aims to guide researchers in selecting suitable network architectures and medical imaging datasets for their specific applications while offering insights into potential solutions for overcoming common hurdles. Through this discussion, we hope to contribute to the advancement of U-Net-based models in medical image segmentation, ultimately improving their practical application in clinical settings.

\bibliographystyle{model1-num-names} 
\bibliography{cas-refs}

\end{document}